\documentstyle[aps,prb]{revtex} 

\begin{document}
\def\be{\begin{equation}}
\def\ee{\end{equation}}
\def\bea{\begin{eqnarray}}
\def\eea{\end{eqnarray}}
\def\le{low-energy }
\def\ham{Hamiltonian }
\def\eh{effective Hamiltonian }
\def\bearst{\begin{eqnarray*}}
\def\eearst{\end{eqnarray*}}
\def\peleven{\parbox{9cm}}
\def\peffec{\peight{\bearst\eearst}\hfill\peleven}
\def\pspace{\peight{\bearst\eearst}\hfill}
\def\ptwelve{\parbox{12cm}}
\def\peight{\parbox{8mm}}

\ifpreprintsty\else 
\twocolumn[\hsize\textwidth%
\columnwidth\hsize\csname@twocolumnfalse\endcsname 
\fi 
 
\draft 
\preprint{} 

\title{Phase diagram of ferrimagnetic ladders with bond-alternation} 

\author{A.~Langari$^{\text a,c}$, M.~Abolfath$^{\text b,c}$
and M. A.~Martin-Delgado$^{\text d}$}
\address{
$^{\text a}$ 
Max-Planck-Institut f\"ur Physik komplexer Systeme, N\"othnitzer Strasse 38,
D-01187 Dresden, Germany\\ }
\address{
$^{\text b}$Department of Physics and Astronomy, 
University of Oklahoma, Norman, Oklahoma 73019-0225}
\address{
$^{\text c}$Institute for Studies in Theoretical Physics and Mathematics,
Tehran, P.O. Box 19395-5531, Iran}
\address{
$^{\text d}$Departamento de
F\'{\i}sica Te\'orica I, Universidad Complutense. Madrid, Spain.}

\date{\today}
\maketitle

\begin{abstract}
\leftskip 2cm 
\rightskip 2cm 

We study the phase diagram of a 2-leg bond-alternation spin-($1/2, 1$) 
ladder for two different configurations using  a quantum renormalization group approach. 
Although  d-dimensional ferrimagnets show gapless behavior, 
we will explicitly show that the effect of 
the spin mixing and the bond-alternation 
can open the possibility for observing an energy gap.  
We show that the gapless phases of such systems can be equivalent
to the 1-dimensional half-integer antiferroamgnets, besides the gapless 
ferrimagnetic phases. We therefore propose a 
phase transition 
between these
two gapless phases that can be seen in the parameter space.
\end{abstract}

\ifpreprintsty\else\vskip1pc]\fi 
\narrowtext 

\section{Introduction}
Since the seminal work of Haldane \cite{r1} quantum spin chains have been 
extensively studied as one of the simplest but most typical quantum 
many-body systems. According to Haldane, the 1-dimensional integer-spin 
Heisenberg antiferromagnets have a unique disordered ground state with a finite 
excitation gap, while half-integer antiferromagnets are gapless and critical
(see \cite{r2} and references therein).
The origin of the difference between half-integer and integer spin chains
can be traced back to the topological $"\theta"$ term in the effective 
non-linear $\sigma$-model description of antiferromagnetic  spin chains and
is believed to be due to non-perturbative effects. Haldane's original 
predictions were based on large-spin arguments and although a general 
rigorous proof is still lacking, several theoretical developments have
helped to clarify the situation and there is now strong experimental and
numerical evidence in support of Haldane's claim \cite{r2,r2a,r2b}.
The massive phase of the integer
spin chains, called Haldane phase, has been understood as 
valence-bond-solid (VBS) states proposed by Affleck {\em et al.} \cite{r3},
wherein each spin-$S$ is viewed as a symmetrized product of $2S$ spinors.
However bond-alternation may drastically change the low-energy behavior
of these systems and
produce phase transitions. 
For instance a bond-alternation
spin-$1/2$ antiferromagnet chain has a finite energy gap as opposed to its 
uniform counterpart \cite{group1,r1213}.
It has been also shown that the phase diagram of the $S=1$ chain 
decomposes into different 
phases by adding bond-alternation \cite{bas1}.
   
Yet another challenge in this area has led to synthesizing 
quasi-one-dimensional bimetallic molecular magnets. 
The search for molecular ferromagnet has led to the discovery of many
interesting molecular magnetic systems. In recent years, quai-one-simensional 
bimettalic molecular magnets, with each unit cell containing two spins
of different spin value have been synthesized \cite{synth}.
These systems contain two transition metal ions per unit cell and have
the general formula ACu(PbaOH)(H$_2$O)$_3$.$2$H$_2$O with
PbaOH=$2$-hydroxo-propylenebis (oxamato) and A=Mn, Fe, Co, Ni. They 
have been characterized as
the alternating (mixed) spin chains or ferrimagnets \cite{r1516}.
It has been shown that one-dimensional ferrimagnets 
have two types of excitations in their low-lying spectrum.
The lowest one has gapless excitations, i.e., they behave like 
ferromagnets with quadratic dispersion relation. The other one which 
is separated by an excitaion gap from the primer one, has antiferromagnetic
behaviour with linear dispersion relation \cite{r17}.
More precisely, the effective Hamiltonian for the lowest spectrum of
a 1-dimensional Heisenberg
ferrimagnets is a 1-dimensional Heisenberg ferromagnet 
with $S=|S_1-S_2|$ \cite{r18a}.
A spin wave study also show the similar behavior is 
seen in two and three dimensions \cite{r18a}. 

Another surprising investigation deals with spin ladders which have recently
attracted a considerable 
amount of attention \cite{r18}.  
They consist of coupled one-dimensional chains and may be 
regarded as interpolating truly one and two-dimensional systems. These models
are useful to study the properties of the high-$T_c$ superconductor
materials.
Theoretical studies have suggested that there are two different universality 
classes for the uniform-spin ladders, i.e., the antiferromagnetic spin-$1/2$ 
ladders are gapful or gapless, 
depending on whether $n_l$ (the number of legs) 
is even or odd \cite{r18}. 
These predictions has been confirmed experimentally by
compounds like ${\rm Sr Cu_2 O_3}$ and ${\rm Sr_2 Cu_3 O_5}$.
However, again bond-alternation changes this universality.
It has been shown that a gapless line which 
depends on the staggered bond-alternation (SBA) parameter $\gamma$, divides 
the gapful phase of a $2$-leg antiferromagnetic spin-$1/2$
ladder into two different phases \cite{r24}, \cite{r24b}. 
Moreover, there are some other configurations, like
the columnar bond-alternation (CBA) [see Fig. \ref{Fig1}(a)] that
introduces new phases for the antiferromagnetic ladders  \cite{r25}.
The appearance of the magnetization plateaus for both chains 
\cite{r2122} and ladders \cite{r23} and the appearance of the new phases for
spin ladders \cite{koga}, 
are also some of the consequences of the bond-alternations.

In this paper, we study the {\it bond-alternation ferrimagnetic ladders}
(BAFL). This model contains a rich  phase diagram.
By making use of a quantum renormalization group (QRG) \cite{r26}, 
we demonstrate that the combination of both spin-mixing and bond-alternation
may open the possibility of observing the
energy `gap' in the ground state of these systems.
We obtain the phase diagram of BAFL in both SBA and CBA configurations.
The SBA configuration consists of a gapfull and two gapless phases, while the
CBS one is composed of two gapless phases. The boundary of these phases where
phase transition takes place is calculated by QRG method.
By using QRG we have obtained the effective Hamiltonian in both strong and 
weak coupling limit of BAFL. We have also shown the structure continued in
the intermediate region.

The paper is organized as follows. In Section II, we present the 
effective Hmiltonian of CBA configuration by QRG method. We show that the 
phase transition occures in the $J'<0$ region. In Sec. III, we then present 
the SBA aconfiguration of ferrimagnetic ladder and discuss the structure of 
phase diagram in both positive and negative $J'$ region.
Sec. IV. is devoted to conclusions.

\section{Columnar bond alternation}

Our model consists of mixed spins $\tau=1/2$ and $S=1$ on a $2$-leg ladder 
(see Fig. \ref{Fig1}). 
The \ham for this system can be divided into three parts: 
$H=H_1^l+H_2^l+H_r$ where $H_\mu^l~ (\mu=1,2)$, and $H_r$ are the  
exchange interaction of the spins inside the $\mu$th leg, 
and the interaction between different legs, as shown in Fig. \ref{Fig1}.
The explicit form of the Hamiltonian is
\bea
\label{hamiltonian}
H_1^l&=&J \sum_{i=1}^{N/2} [(1+\gamma)\vec \tau_1(2i-1) \cdot \vec S_1(2i) \nonumber \\
&+&(1-\gamma)\vec S_1(2i) \cdot \vec \tau_1(2i+1)] \nonumber \\
H_2^l&=&J \sum_{i=1}^{N/2} [(1+\gamma)\vec S_2(2i-1) \cdot \vec \tau_2(2i) \nonumber \\
&+&(1-\gamma)\vec \tau_2(2i) \cdot \vec S_2(2i+1)] \nonumber \\
H_r&=&J' \sum_{i=1}^{N/2} [\vec \tau_1(2i-1) \cdot \vec S_2(2i-1) \nonumber \\
&+&\vec S_1(2i) \cdot \vec \tau_2(2i)] 
\eea

The ladder contains $N$ rungs and by assuming the periodic boundary condition
for each leg, we identify the first rung to the $N+1$-th rung. 
Through out this paper we assume $J$ is positive 
but $J'$ can be positive or negative as a tunable parameter. 
Note that for the bond-alternation parameter, 
$\gamma \rightarrow -\gamma$ amounts to sublattice exchange $n \rightarrow
n+1$, and therefore we consider $0\leq\gamma\leq1$.

By QRG, we divide the \ham  
into intra-block ($H^B$) and inter-block ($H^{BB}$) parts. After
diagonalizing the first part, a number of \le eigenstates are kept to 
project the full \ham onto the renormalized one. 
As opposed to other powerful
techniques, such as the density matrix renormalization group (DMRG) \cite{r27},
the QRG approach is much less complicated yielding analytical
 results, although QRG does not give as accurate numerical
results as DMRG, nevertheless
 its simplicity can give a good qualitative picture
of the phase diagram \cite{r28}.
In this paper four interesting configurations are in order:
positive and negative value of $J'$ for both CBA and SBA configurations.

\subsection{The case $J'>0$}

The CBA configuration [Fig. \ref{Fig1}(a)] and $J' > 0$.
Let us first consider the strong 
coupling limit ($J' \gg J$). 
Since the interaction between two sites on each rung is strong, then
each rung can be considered as
the isolated block in the first step of the QRG, i.e., $H^B=H_r$.
The Hilbert space of each block ($\tau-S$) consists of two multiplets
whose total spin are $3/2$ and $1/2$. The corresponding energies 
for these two configurations are $J'/2$ and $-J'$. 
Therefore we keep the $S=1/2$ multiplet as the basis 
for constructing the embedding operator $T$ to project the 
full \ham onto the truncated 
Hilbert space ($H_{\rm eff}=T^\dagger H T$) \cite{r29}. 
Finally the \eh (in which each rung is mapped to a single
site) can be obtained
\bea
\label{l2}
H_{1{\rm eff}}= &-&NJ'-\frac{8}{9} J \sum _{i=1}^{N/2}
[(1+\gamma)\vec S'(2i-1) \cdot \vec S'(2i) \nonumber \\
&+&(1-\gamma)\vec S'(2i) \cdot \vec S'(2i+1)],
\eea
where $S'=1/2$ and $i$ is the label of the
sites on new chain which represents the $i$th rung of the original ladder.
Eq. (\ref{l2}) is the Hamiltonian of a spin-$1/2$ (bond-alternation) 
ferromagnetic chain. 
It exhibits gapless excitations as well as the ferromagnetic ground state.
Equivalently, the original ladder exhibits 
the ferrimagnetic ground state, in the sense that both 
magnetization ($m=<\tau>+<S>=0.5$) and the staggered one 
($sm=<S>-<\tau>=5/6$) are non-zero with a gapless excitations.

Now, let us consider the weak coupling limit ($J' \ll J$). In this case the
stronger bonds, e.g. $J(1+\gamma)$, appear on the legs. 
They are considered as the isolated blocks. 
The other bonds (the weaker ones) are considered as $H^{BB}$. The QRG
procedure leads to the Heisenberg (spin-$1/2$) ferromagnet ladder.  
In this regime the spectrum of the system is similar to the strong coupling 
limit. 

For the intermediate region where $J'\simeq J$, 
the two types of blocking which has been considered in
the strong and weak coupling limits seems to be not suitable.
In this regard we have considered a 4-sites block which consists of 
both rung and leg interactions [Fig.\ref{Fig2r} (a)]. 
Decomposition of ladder to 4-sites blocks leads to two types of 
 building blocks,
which is shown in [Fig.\ref{Fig2r} (b)] and [Fig.\ref{Fig2r} (c)].
The lowest energy multiplet of the Hamiltonian in [Fig.\ref{Fig2r} (b)]
has total spin $\hat S=5/2$ and the corresponding one in [Fig.\ref{Fig2r} (c)]
has $\tau'=1/2$. The embedding operator is constructed from these two 
multiplets and finally the effective Hamiltonian ($H_{\rm eff}=T^\dagger H T$)
is a $(1/2,5/2)$ ferrimagnetic chain.
\bea
\label{hi1}
H_{i-eff}=-\frac{9}{8}NJ&+&\frac{40}{63}J\sum_{i=1}^{N/2}
[\vec \tau'(2i-1) \cdot \vec{\hat S}(2i) \nonumber \\ 
&+&\vec{\hat S}(2i) \cdot \vec \tau'(2i+1)]
\eea
In the next step of QRG procedure the Hamiltonian in Eq.(\ref{hi1}) is 
projected to a $\tilde S=2$ ferromagnetic Heisenberg chain \cite{r18a}.
\be
\label{hi2}
H_{eff}(\tilde S=2)=-\frac{9}{8}J+\frac{40}{63}J\sum_{i=1}^{N/4}
\vec{\tilde S}(i)+\vec{\tilde S}(i+1)
\ee
Thus the magnetization ($m$) and staggered magnetization ($sm$) of the 
original ladder in the intermediate regime are : $m\simeq 0.2292 \;;\;
sm\simeq 0.4514$, which shows ferrimagnetic order.          
We therefore find the different regimes have similar structure
(the ferrimagnetic ground state with the gapless excitations)
as long as $J'$ is positive.

\subsection{The case $J'<0$}
This configuration has more interesting 
features.
Let us first consider the strong coupling limit 
where $H^B$ is constructed by $H_r$. Since $J'$ is negative the 
\le multiplet has total spin $3/2$. Thus this subspace is considered as 
the effective Hilbert space for the first step of the QRG.
The \eh ($H_{2{\rm eff}}$) can be obtained by
projecting each operator onto the effective Hilbert space
\bea
\label{l3}
H_{2{\rm eff}}&=&\frac{NJ'}{2}+\frac{4}{9}J \sum_{i=1}^{N/2}
[(1+\gamma) \vec S''(2i-1) \cdot \vec S''(2i) \nonumber \\
&+& (1-\gamma) \vec S''(2i) \cdot \vec S''(2i+1)]
\eea
where $S''$ is a spin $3/2$. Hamiltonian (\ref{l3}) is a 1-dimensional
Heisenberg spin-$3/2$ antiferromagnet with alternating bonds.
It is known \cite{r1213} that this
model is gapful when $\gamma > \gamma_c$ and gapless otherwise. 
Although the one-, two- and three-dimensional
ferrimagnets show gapless behavior,
 the combinations of the  
spin-mixing and bond-alternations yields the possibility for developing of 
an energy gap. 
This can be compared to the spin-$1/2$ CBA Heisenberg ladders \cite{r25} where 
the model is gapful in the whole range of parameter space ($J, J', \gamma$)
except on a critical line, in the region where $J'$ is negative.
In the next step of QRG, $H_{2{\rm eff}}$ will be projected to a spin-$1/2$
XXZ model in the presence of an external magnetic field ($h$). It is known
\cite{r31} that XXZ$+h$ model has the critical line $h_c=\Delta+1$,
where $\Delta$ is the 
anisotropy in the $\hat{z}$-direction \cite{r28}.
If $h>h_c$ ($h<h_c$) the model is gapful (gapless) and
the value of gap is proportional to the strength of magnetic field. 
We find that 
$\gamma_c=3/7 (\simeq 0.428)$ for our model which is 
close to the
DMRG results $(\simeq 0.42)$ for spin-$3/2$ dimerized chain \cite{r1213}.
Therefore in the ferromagnetic region ($J'<0$) and strong coupling 
($|J'|\gg J$)  there is a critical value for $\gamma$ below which 
($\gamma<\gamma_c$) the BAFL is gapless and its ground state has
quasi-long range order (quasi-LRO)
where both $m$ and $sm$ are zero
and correlations decay algebraically.
This is equivalent to the uniform
1-dimensional spin-$3/2$ antiferromagnets. For $\gamma>\gamma_c$ the model is gapful.

In the weak coupling limit ($|J'| \ll  J$), 
the strongest bonds, e.g. $J(1+\gamma)$, 
of the ladder are considered as building blocks of QRG, 
and the remaining bonds are treated as $H^{BB}$.
The \eh in this case is
\bea
\label{l5}
H_{3{\rm eff}} &=&
\frac{8|J'|}{9} \sum_{i=1}^{N/2} \vec S'_1(i) \cdot \vec S'_2(i) \nonumber \\
&-&\frac{4J(1-\gamma)}{9} \sum_{i=1}^{N/2} \sum_{\mu=1,2}                    
\vec S'_{\mu}(i) \cdot \vec S'_{\mu}(i+1)
\eea
Here $S'=1/2$ and we have neglected a constant term. 
The effective Hamiltonian $(H_{3{\rm eff}})$ is a `double-layer' model of spin-$1/2$.
Neglecting the inter-site terms (for a moment), 
each couple of sites has a multiplet
of states with total angular momentum $\ell = 0, 1$ \cite{note1}, where
each pair of 
inter-layer spins on the two layers, behaves like a single
quantum rotor where ${\vec L} = {\vec S_1} + {\vec S_2}$ and
${\vec n} = ({\vec S_1} - {\vec S_2})/2S$. 
Considering the 
intra-layer terms, one may map the Hamiltonian (\ref{l5})
to that 1-dimensional quantum rotor model \cite{Senthil}
\bea
\label{l5_1}
H_{3{\rm eff}} = \frac{g}{2} \sum_{i=1}^{N/2} {\vec L}^2_i 
- K \sum_{i=1}^{N/2}({\vec n}_i\cdot{\vec n}_{i+1} + 
{\vec L}_i\cdot{\vec L}_{i+1}),
\eea
where $g\equiv 16|J'|/9$ and $K=4J(1-\gamma)/9$. The mean field phase
diagram of this model is governed by the gapped quantum paramagnet
when $\gamma \rightarrow 1$ and the partially polarized ferromagnet
when $\gamma \rightarrow 0$.
The dominant term of the first limit is the antiferromagnetic exchange term
along the rungs which makes singlets as the base structure for ground state of the
renormalized ladder. Thus the ground state is unique and has a finite energy gap
to the first excited state. This gapful phase occurs in the region
where $|J'|/J \gg 1-\gamma$, which is the continuation 
of the gapful phase in the strong coupling limit.
Thus we have no-LRO in the ground state of the original ladder and finite gap
in this region.
For the latter limit, there is a competition between the ferromagnetic
term and the antiferromagnetic term in ($\ref{l5}$).
Note that the dominant term in $H_{3{\rm eff}}$ is the ferromagnetic interaction 
along the legs. 
Then the ground state is composed of two ferromagnetic 
ordered chains which 
are aligned oppositely due to antiferromagnetic interaction 
($8|J'|/9$). But this classical antiferromagnetic alignment
fluctuates along the $\hat{z}$-axis and the magnetization is reduced 
along this direction.

In the intermediate region (where $|J'| \simeq J$), the ladder 
is decomposed to 4-sites plaquettes if $\gamma\rightarrow 1$. 
Note that $S_z = 0$ is 
the unique ground state of any 4-site plaquette at 
$2J=-J'=1$
(one may see this after diagonalizing the Hamiltonian).
But every plaquette is on the $S_z = 0$ state, since the ladder is
disconnected.
This ground state is 
not
degenerate and disorder (with no-LRO) and
there is a finite energy gap to the first excited state.
In other words the whole gapful phase of
the ladder is in the VBS phase.

On the other extreme case, when $\gamma$ is negligible,
the gapless phase behaves differently in long wave-length limit.
Let us first suppose $|J'|>J$. In this case, 
each rung behaves as a spin-$3/2$ (after the first step of the QRG),
since the spins on the same rung are coupled by the ferromagnetic 
interaction ($J'<0$). 
Hence the whole ladder is identified by a Heisenberg spin-$3/2$ 
antiferromagntic chain. The ground state is gapless with 
quasi-LRO,
and the correlation functions falls off algebraically (the correlation length
$\xi$, is infinite).
But if $|J'|<J$ the ladder can be 
considered as
two quantum ferrimagnetic 
chains that interact via weak ferromagnetic coupling (through their rungs). 
The correlation length is small and the correlation
functions falls off exponentially.
One may naturally expect that at $|J'| \sim J$
a phase transition is observed between two gapless phases. 
One phase is the half-integer quantum antiferromagnets 
(when $|J'|\gg J$) with $\xi = \infty$ and 
quasi-LRO
and another phase is the ferrimagnetic phase 
(when $|J'|\ll J$) with $\xi \sim a$ ($a$ is the 
lattice spacing). 
The order parameter to specify this phase transition is ${\tilde m}=|m_1-m_2|$,
where $m_{1(2)}$ is the magnetization per site of the 1-st (2-nd) leg of 
ladder. ${\tilde m}$ is zero where $(J'/J)<-1$ and ${\tilde m}=0.5$
for $(J'/J)>-1$. The total magnetization (m) and staggered one (sm) are
zero on both side of this critical line.
This completes the phase diagram of the CBA configuration. It is 
depicted in Fig. \ref{Fig2}(a).


\section{staggered bond alternation}
\subsection{The case $J'>0$}
The SBA configuration is shown ind [Fig. \ref{Fig1}(b)]. 
In this region 
the effective Hamiltonian shows ferromagnetic behavior [similar to the 
CBA ($J'>0$) configuration ] and the ladder behaves like
the gapless ferrimagnets. 
As an example, we may apply the ``snake mechanism" of reference \cite{r24}:
choosing $\gamma=1$ and $J'=2J$ the ladder degenerates into a uniform
alternating spin-$1/2-1$ ferrimagnetic chain which has gapless excitations.
More precisely the effective Hamiltonian ($H_{4eff}$) in the strong
coupling limit ($J' \gg J$) is
\be
\label{h4eff}
H_{4eff}=-NJ'-\frac{8}{9}J\sum_{i=1}^{N} \vec S'(i) \cdot \vec S'(i+1)
\ee
where $S'=1/2$ and the block Hamiltonian for QRG procedure is considered to 
be $H_r$. In the weak coupling limit ($J'\ll J$) where the strongest bonds are
$J(1+\gamma)$, the QRG procedure leads to a strip of triangular lattice
as the effective Hamiltonian ($H_{5eff}$).
\bea
\label{h5eff}
H_{5eff}&=&-NJ(1+\gamma)-\frac{4}{9}J(1-\gamma)\sum_{\mu=1}^2 
\sum_{i=1}^{N} \vec S'_{\mu}(i) \cdot \vec S'_{\mu}(i+1)  \nonumber \\
&-&\frac{4}{9}J'\sum_{i=1}^{N}[\vec S'_1(i) \cdot \vec S'_2(i) + 
\vec S'_2(i) \cdot \vec S'_1(i+1)]
\eea
The effective Hamiltonians ($H_{4eff}\;,\;H_{5eff}$) are $S=1/2$
ferromagnetic Heisenberg model. Thus in both of these cases the model has
gapless excitations. Since both of $m$ and $sm$ are not zero, we have
ferrimagnetic order in the whole part of ($J'>0$) region.

\subsection{The case $J'<0$}

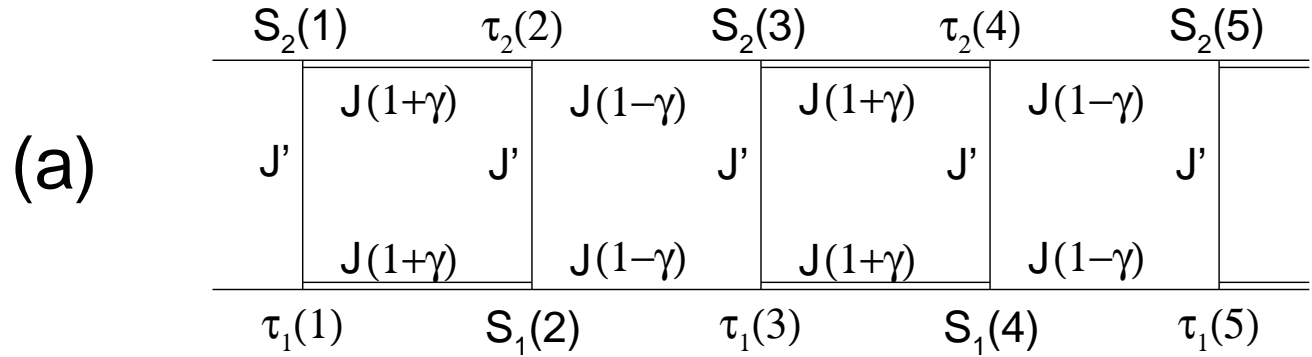
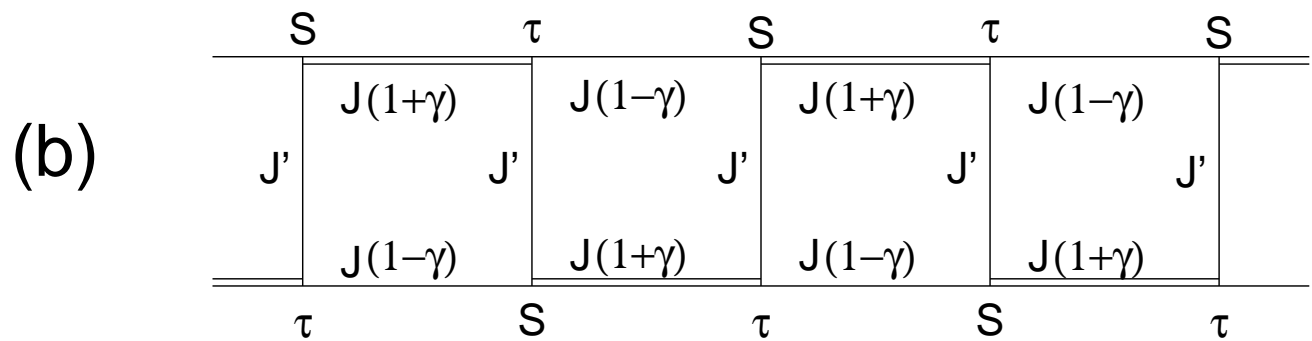
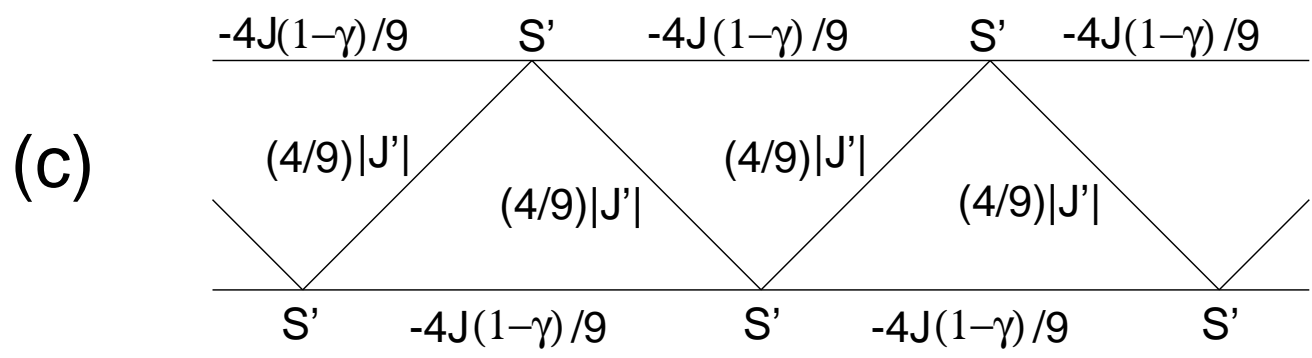
\begin{figure}

\caption{The ladder realization $\tau=1/2$ and $S=1$, 
(a) Columnar bond-alternation (CBA), (b) staggered bond-alternation (SBA), 
(c) schematic illustration of effective Hamiltonian in SBA 
configuration at $|J'|\ll J$, (where $S'=1/2$).}
\label{Fig1}
\end{figure}

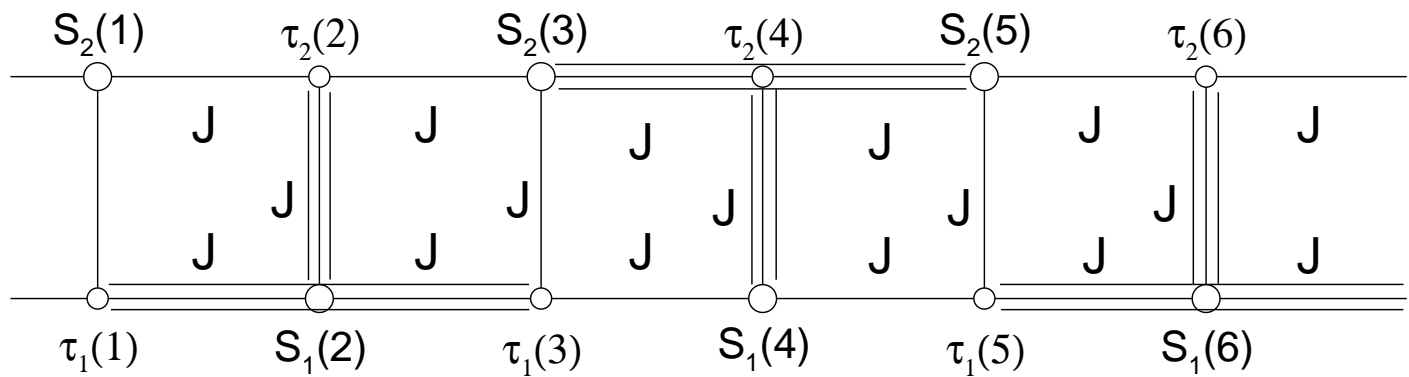
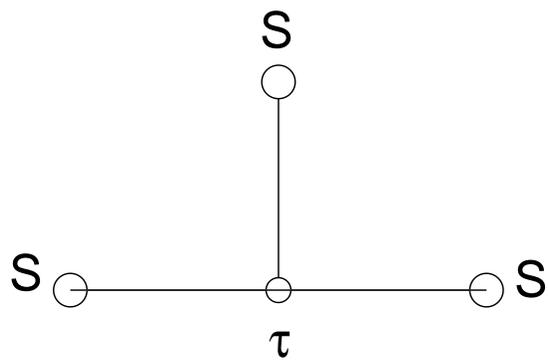
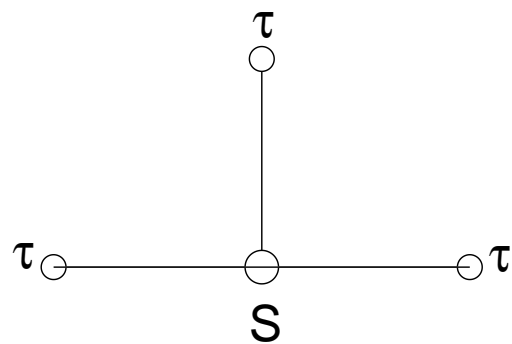
\begin{figure}
\caption{(a) Decomposition of ladder into 4-sites blocks in the intermediate
region ($J \sim J'$), (b) three spin-$1$ and a spin-$1/2$, 
(c) three spin-$1/2$ and a spin-$1$}

\label{Fig2r}
\end{figure}

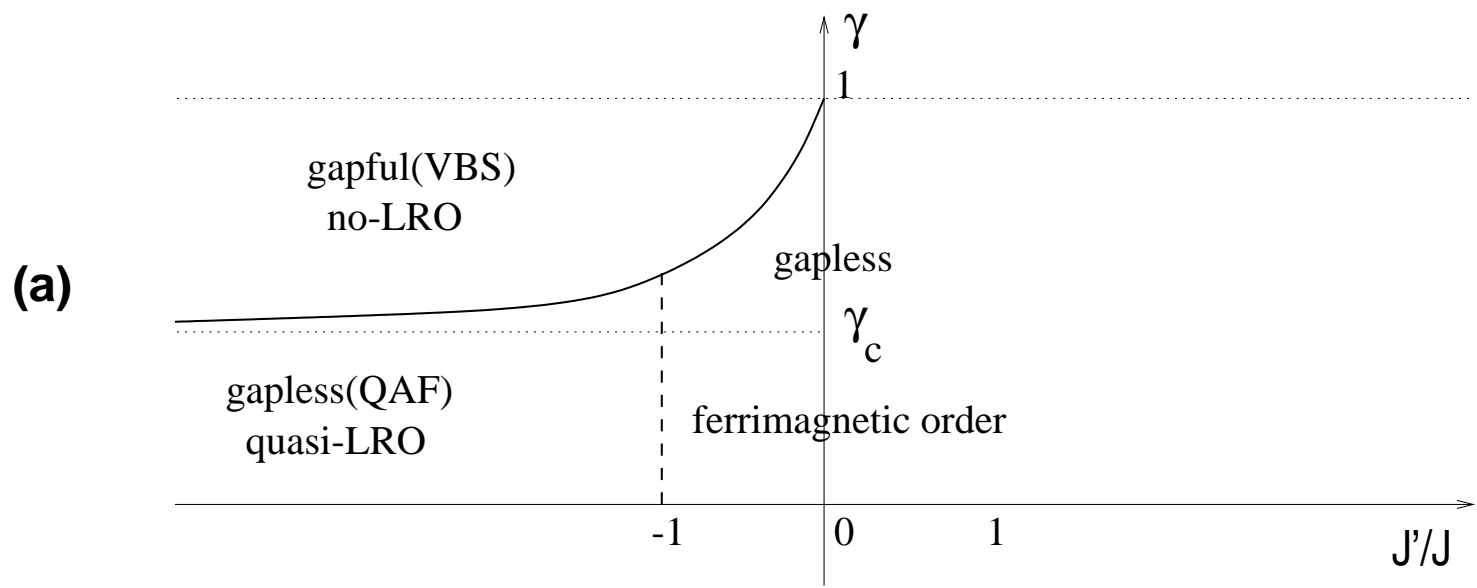
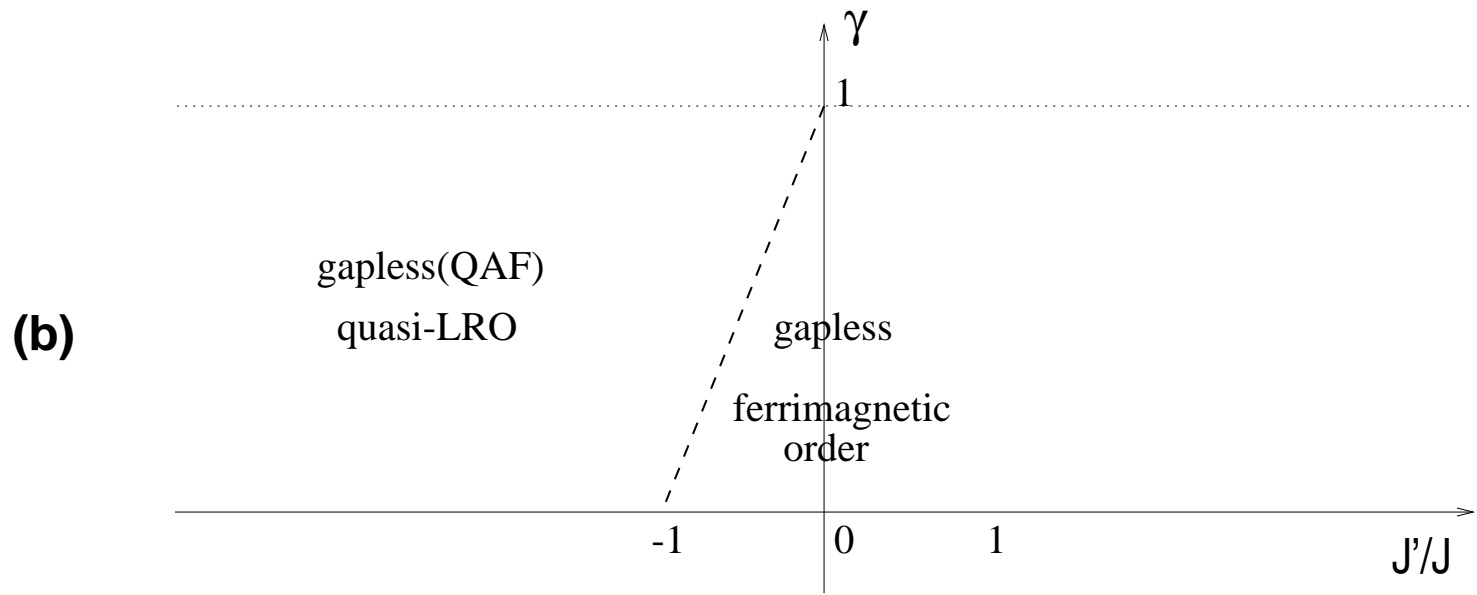
\begin{figure}
\caption{Phase digram of bond-alternation ferrimagnetic $2$-leg ladder,
(a) Columnar bond-alternation (CBA), (b) staggered bond-alternation (SBA),
solid line is the critical line 
between gapless and gapful phases and the dashed one shows the
critical line between two gapless phases.} 
\label{Fig2}
\end{figure}

In this region and at the strong coupling limit ($|J'| \gg J$) the 
ladder is mapped to a  
uniform Heisenberg spin-3/2 antiferromagnetic chain, where the alternation
parameter is disappeared at the first step of the QRG. 
Thus the whole range of $\gamma~ (\in [0, 1])$ is gapless and 
disordered. 
The system exhibits quasi-LRO
with $\xi = \infty$.
But when $|J'| \ll J$, the \eh is equivalent to a two 
1-dimensional spin-1/2 ferromagnetic
Heisenberg chains. These two chains interact by an antiferromagnetic 
coupling on a triangular ladder as shown in Fig. \ref{Fig1}(c). 
At $\gamma=1$ the ladder transforms to a 1-dimensional spin-1/2
antiferromagnetic chain which is gapless with 
quasi-LRO.
But at $\gamma=0$ (where the CBA is equivalent to SBA) 
the ladder is equivalent to two ferromagnetic chain
which are coupled antiferromagnetically. 
It represents a ferrimagnetic phase where $\xi \sim a$. 
As we have illustrated above the competition between the
coupling constants in Fig. \ref{Fig1}(c) leads to one of the above extreme
cases. 
In other words, if $|J'|/J > 1-\gamma$ 
the system is equivalent to the Heisenberg
spin-1/2 antiferromagnetic chains. 
For another opposite limit, the system is equivalent to
the Heisenberg ferrimagnetic chains. The dashed line in Fig. 2(b) 
which separates these two phases, represents the 
critical line.

\section{conclusion}

In summary, we have obtained the phase digram ($J'/J, \gamma$) of a 2-leg 
bond-alternating ($1/2, 1$) ferrimagnetic ladder. In CBA configuration, 
Fig. 1(a), there
exists a gapful (VBS) phase which is separated from two different gapless
phases by the critical lines. The phase transition 
between gapless phases takes place when 
$J'/J \sim -1$.
For the SBA configuration, Fig.1(b), the whole phase diagram is gapless.
One phase contains the ferrimagnetic behavior and another one is
equivalent to the Heisenberg spin-$1/2$ antiferromagnetic chains.
In the latter region, the correlation function falls off algebraically.
The transition
between these two gapless phases takes place when
$J'/J \sim (\gamma-1)$.

Although the ($1/2, 1$) ferrimagnetic system is condsidered as 
a generic model for all ($S_1, S_2$) systems, we have found
that this is  not  longer true 
in ladders
if one considers the bond-alternation effects, where both
spin-mixing and bond-alternation may change the quantum
phase transitions. The dependence on different spins and the number of legs
should be considered in future investigations. 

\acknowledgments
We would like to thank  H. Hamidian, Hsiu-Hau Lin and G. Sierra
for helpful
discussions and useful comments.
M.A. acknowledges support from EPS-9720651 and a grant from the Oklahoma
State Regents for Higher Education.
M.A.M.-D. is  supported by the DGES spanish grant
PB97-1190.

\begin{references}
\bibitem{note1}
For the low energy properties of interest in this paper, the upper
bound do not play an important role.
\end{references}
\end{document}